\documentclass[]{aa}

\usepackage{txfonts}
\usepackage[colorlinks, breaklinks]{hyperref}
\hypersetup{linkcolor=blue, citecolor=blue, filecolor=black, urlcolor=blue}

\newcommand{\orcid}[1]{
\href{https://orcid.org/#1}{
\includegraphics[height=11pt]{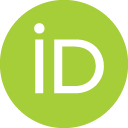}}}

\newcommand{\prob}[2]{\mathcal{P}\left( #1 \mid #2\right)}

\begin{document}

\title{Redshift evolution of the underlying type Ia supernova stretch
distribution}

\titlerunning{Redshift evolution of the underlying type Ia supernova stretch
distribution}
\authorrunning{N.~Nicolas et al.}

\author{
    N. Nicolas \thanks{n.nicolas@ip2i.in2p3.fr, equal contribution} \inst{1} 
    \and M. Rigault \thanks{m.rigault@ip2i.in2p3.fr, equal contribution} \inst{1}
    \orcid{0000-0002-8121-2560}
    \and\\ Y. Copin \inst{1}
    \orcid{0000-0002-5317-7518}
    \and R. Graziani \inst{2}
    \and G. Aldering\inst{3}
    \and M. Briday\inst{1}
    \and Y.-L. Kim\inst{1}
    \orcid{0000-0002-1031-0796}
    \and J. Nordin\inst{4}
    \and Saul Perlmutter\inst{3}
    \and M. Smith\inst{1,5}
    \orcid{0000-0002-3321-1432}
}

\institute{Univ Lyon, Univ Claude Bernard Lyon 1, CNRS, IP2I Lyon / IN2P3, IMR
    5822, F-69622, Villeurbanne, France
    \and 
    Université Clermont Auvergne, CNRS/IN2P3, Laboratoire de
    Physique de Clermont, F-63000 Clermont-Ferrand, France.
    \and
    Physics Division, Lawrence Berkeley National Laboratory, 
    1 Cyclotron Road, Berkeley, CA, 94720 
    \and
    Institut fur Physik, Humboldt-Universität zu Berlin, Newtonstr. 15,
    12489 Berlin
    \and
    University of Southampton: Southampton, GB
}

\date{Submitted to A\&A\ the 19th of May 2020}

\abstract{The detailed nature of type Ia supernovae (SNe~Ia) remains uncertain,
    and as survey statistics increase, the question of astrophysical systematic
    uncertainties arises, notably that of the evolution of SN~Ia populations. We
    study the dependence on redshift of the SN~Ia \texttt{SALT2.4} light-curve
    stretch, which is a purely intrinsic SN property, to probe its potential
    redshift drift. The SN stretch has been shown to be strongly correlated with
    the SN environment, notably with stellar age tracers. We modeled the
    underlying stretch distribution as a function of redshift, using the
    evolution of the fraction of young and old SNe~Ia as predicted using the
    SNfactory dataset, and assuming a constant underlying stretch distribution
    for each age population consisting of Gaussian mixtures. We tested our
    prediction against published samples that were cut to have marginal
    magnitude selection effects, so that any observed change is indeed
    astrophysical and not observational in origin. In this first study, there
    are indications that the underlying SN~Ia stretch distribution evolves as a
    function of redshift, and that the age drifting model is a better
    description of the data than any time-constant model, including the
    sample-based asymmetric distributions that are often used to correct
    Malmquist bias at a significance higher than 5 $\sigma$. The favored
    underlying stretch model is a bimodal one, composed of a high-stretch mode
    shared by both young and old environments, and a low-stretch mode that is
    exclusive to old environments. The precise effect of the redshift evolution
    of the intrinsic properties of a SN Ia population on cosmology remains to be
    studied. The astrophysical drift of the SN stretch distribution does affect
    current Malmquist bias corrections, however, and thereby the distances that
    are derived based on SN that are affected by observational selection
effects. We highlight that this bias will increase with surveys covering
increasingly larger redshift ranges, which is particularly important for the
Large Synoptic Survey Telescope.}

\keywords{Cosmology -- Type Ia Supernova -- Systematic uncertainties}
\maketitle

\section{Introduction}

Type Ia supernovae (SNe Ia) are powerful cosmological distance indicators that
enabled the discovery that the Universe's expansion accelerates
\citep{riess1998, perlmutter1999}. They remain a key cosmological probe today
for understanding the properties of dark energy (DE) because they are the only
tool that can precisely map the recent expansion rate ($z<0.5$) when DE is
driving it \citep[e.g.,][]{scolnicastro2020}. Type Ia supernovae are also the
key to directly measuring the Hubble constant ($H_0$) if their absolute
magnitude can be calibrated \citep{riess2016, freedman2019}. Interestingly, the
value of $H_0$ that is derived when the SNe~Ia are anchored to Cepheids
\citep[the Supernovae, $H_0$, for the Equation of State of dark energy
project,][]{riess2009, riess2016} is $\sim5\sigma$ higher than what is predicted
from cosmic microwave background (CMB) data measured by Planck assuming the
standard $\Lambda$CDM \citep{planck2018, riess2019, reid2019}, or when the SN
luminosity is anchored at intermediate redshift by the baryon acoustic
oscillation (BAO) scale \citep{feeney2019}. While using the tip of the red giant
branch technique in place of the Cepheids seems to favor an intermediate value
of $H_0$ \citep{freedman2019, freedman2020}, time-delay measurements from strong
lensing also appear to favor high $H_0$ values \citep{wong2019}.

The $H_0$ conundrum has received much attention because it could be a sign of
new fundamental physics. No simple solution is so far able to accommodate this
$H_0$ conundrum when all other probes are accounted for, however
\citep{knox2019}. Alternatively, systematic effects affecting one or several of
the aforementioned analyses might explain at least some of this discrepancy.
\cite{rigault2015} suggested that SNe Ia from the Cepheid calibrator sample
differ by construction from those in the Hubble flow sample, as the former
strongly favors young stellar populations while the latter does not. This
selection effect would affect the derivation of $H_0$ if SNe~Ia from young and
older environments differed in average standardized magnitudes. 

The relation between SNe~Ia and their host galaxy properties has been studied
extensively. The first key finding was that the standardized SNe~Ia magnitudes
significantly depend on the host galaxy stellar mass, with SNe~Ia from high-mass
host galaxies being brighter on average \cite[e.g.,][]{kelly2010, sullivan2010,
childress2013, betoule2014, kim19, smith2020}. This mass-step correction is
currently used in cosmological analyses \citep[e.g.,][]{betoule2014,
scolnic2018a}, including to derive $H_0$ \citep{riess2016, riess2019}. The
underlying connection between the SNe and their host galaxies remains unclear,
however, particularly when global properties such as the host stellar mass are
used. This in turn raises the question of the accuracy of such corrections
because the global properties of the host galaxies evolve with redshift. More
recently, studies have used the local SN host galaxy environment to probe more
direct connections between the SNe and their host galaxy environments
\citep{rigault2013}, showing that local age tracers such as the local specific
star formation rate (LsSFR) or the local color are more strongly correlated with
the standardized SN magnitude \citep{roman2018, kim18, rigault2020}. The
identification of SN~Ia spectral features that are correlated with LsSFR
\citep{nordin2018} further support this connection. These results suggest age as
the driving parameter underlying the mass step. If this is true, it would have a
significant effect for cosmology because the environmental correction that would
need to be applied to SN standardization could strongly vary with redshift
\citep{rigault2013, childress2014, scolnic2018a}. The importance of local SN
environmental studies remains highly debated, however \cite[e.g.,][]{jones2015,
jones2019}, and especially the effect of such an astrophysical bias has on the
derivation of $H_0$ \citep{jones2015, riess2016, riess2018, rose2019}. 

The concept of the SN~Ia age dichotomy arose with the study of the SN~Ia rate.
\cite{mannucci2005}, \cite{scannapieco2005}, \cite{sullivan2006} and
\cite{smith2012} have shown that the relative SNe~Ia rate in galaxies could be
explained if two populations existed, a young population that follows the host
star formation activity, and an old population that followed the host stellar
mass (the so-called prompt-and-delayed or A+B model). \cite{rigault2020} used
the LsSFR to determine the younger (those with a high LsSFR) and the older
(those with a low LsSFR) populations. Furthermore, since the first SNe~Ia host
analyses, the SN stretch has been known to be strongly correlated with the SN
host galaxy properties \citep{hamuy1996, hamuy2000}. This correlation has been
extensively confirmed since then \citep[e.g.,][]{neill2009, lampeitl2010,
gupta2011, dandrea2011, pan2014}. Following the A+B model and the connection
between SN stretch and host galaxy properties, \cite{howell2007} first discussed
the potential redshift drift of the SN stretch distribution. In this paper we
revisit this question using the most recent SNe~Ia datasets.

We here step aside from the cosmological analyses to probe the validity of our
modeling of the SN population, which we claim is constituted of two age
populations \citep{rigault2013, rigault2015, rigault2020}: an older and a
younger population, the former having lower light-curve stretches on average and
being brighter after standardization. We use the correlation between the SN age
as probed by the LsSFR and the SN stretch to model the expected evolution of the
underlying SN stretch distribution as a function of redshift. This modeling
relies on three assumptions: (1) There are two distinct populations of SNe~Ia,
(2) the relative fraction of each of these populations as a function of redshift
follows the model presented in \cite{rigault2020}, and (3) the underlying
distribution of stretch for each age sample is constant. This paper tests this
specific model with datasets from the literature. 

We present the sample we used for this analysis in Section~\ref{sec:sample}. It
was derived from the Pantheon catalog \citep{scolnic2018a}. We discuss the
importance of obtaining a ``complete'' sample, that is, a sample that is\
representative of the true underlying SNe Ia distribution, and how we built one
from the Pantheon sample. We then present our modeling of the distribution of
stretch in Section~\ref{sec:modeling} and our results in
Section~\ref{sec:results}. In this section, we test whether the SN stretch
distribution evolves as a function of redshift and determine whether the
aforementioned age model agrees well with this evolution. We briefly discuss
these results in the context of SN cosmology in Section~\ref{sec:discussion},
and we conclude in Section~\ref{sec:ccl}.

\section{Complete sample construction}\label{sec:sample}

The ideal SN~Ia sample for studying this question would be a very deep,
large-area, volume-limited sample. This would capture the true underlying
stretch distribution function, and we would then study how it evolves with
redshift. No such sample exists, therefore we must first construct subsamples
from existing high-redshift SN~Ia samples that are as near to volume-limited as
possible.

\subsection{Applying redshift cuts}\label{ssec:cuts}

We based our analysis on the most recent comprehensive SNe~Ia compilation, the
Pantheon catalog from \cite{scolnic2018a}. A naive approach to testing the SN
stretch redshift drift would be to simply compare the observed SN stretch
distributions in a few redshift bins. In practice, however, differential
observational selection effects will affect the observed SN stretch
distributions. Because the observed SN~Ia magnitude correlates with the
light-curve stretch (and color), the first SNe~Ia that a magnitude-limited
survey will miss are those at the lowest stretch (which are the reddest).
Consequently, if magnitude-related observational selection effects are not
accounted for, true population drift might be confused with survey properties,
and vice versa.

Assuming sufficient (and unbiased) spectroscopic follow-up for acquiring SN
types and host galaxy redshifts, the observation selection effects of
magnitude-limited surveys should be negligible below a given redshift at which
even the faintest normal SNe~Ia can be observed. Aiding in the construction of
nearly volume-limited subsamples is the fact that the SN~Ia population trails
off toward fainter SNe~Ia. A complication is that complete spectroscopic
follow-up has not always been the norm, as discussed below. In contrast,
targeted surveys have highly complex observational selection functions and so
are discarded from our analysis. High-redshift SN cosmology samples, such as the
Pantheon sample, are predominately assembled from magnitude-limited surveys from
which volume-limited SN~Ia subsamples can be constructed.

\begin{figure}
    \centering
    \includegraphics[width=0.95\linewidth]{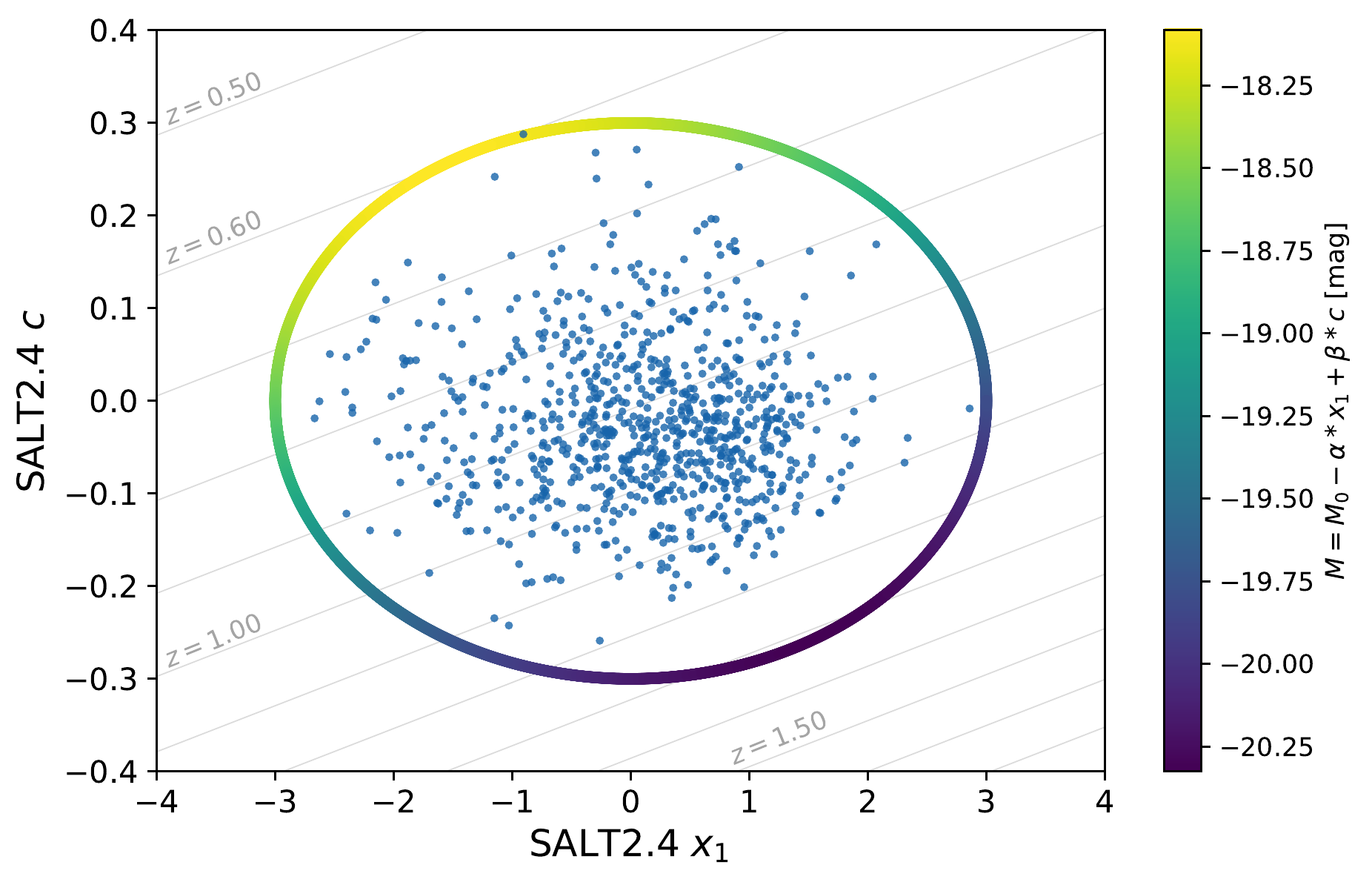}
    \caption{\textsc{\texttt{SALT2.4}} stretch ($x_1$) and color ($c$)
        light-curve parameters of SNe~Ia from the SDSS, PS1, and SNLS samples of
        the Pantheon catalog. The individual SNe are shown as blue dots. The
        ellipse $(x_1=\pm3, c=\pm0.3)$ is displayed, colored by the
        corresponding standardized absolute magnitude using the $\alpha$ and
        $\beta$ coefficients from \cite{scolnic2018a}. The diagonal gray lines
        represent the $(x_1, c)$ evolution for $m = m_{\lim}$ for $z$ between
        $0.50$ and $z=1.70$ using the SNLS $m_{\lim}$ of $24.8$~mag.}
    \label{fig:maglim}
\end{figure}

We present in Fig.~\ref{fig:maglim} the light-curve stretch and color of SNe~Ia
from the following surveys: PanStarrs \citep[PS1][]{rest2014}, the Sloan Digital
Sky Survey \citep[SDSS][]{frieman2008}, and the SuperNovae Legacy Survey
\citep[SNLS][]{astier2006}. An ellipse in the \textsc{\texttt{SALT2.4}} $(x_1,
c)$ plane with $x_1 = \pm 3$ and $c = \pm 0.3$ encapsulates the full parent
distribution \citep{guy2007, betoule2014}; see also \citet{bazin2011} and
\citet{campbell2013} for similar contours, the second using a more conservative
$|c| \leq 0.2$ cut. Assuming the SN absolute magnitude with $x_1=0$ and $c=0$ is
$M_0=-19.36$ \citep{kessler2009,scolnic2014}, we can derive the absolute
standardized magnitude at maximum of light $M = M_0 - \alpha x_1 + \beta c$
within the aforementioned ellipse given the standardization coefficient
$\alpha=0.156$ and $\beta=3.14$ from \cite{scolnic2018a}: The faintest SN~Ia is
that with $(x_1=-1.65, c=0.25)$ and an absolute standardized magnitude at peak
in Bessel $B$ band of $M^{t_0}_{\min} = -18.31$~mag. This object typically ought
to be detected 5 days before and a week after peak to build a suitable
light-curve, the effective limiting standardized absolute magnitude is
approximately $M_{\lim} = -18.00$~mag. Hence, given the magnitude limit
$m_{\lim}$ of a magnitude limited survey, we can derive the maximum redshift
$z_{\lim}$ above which the faintest SNe~Ia will be missed using the relation
between apparent magnitude, redshift, and absolute magnitude $\mu(z_{\lim}) =
m_{\lim} - M_{\lim}$.

We therefore considered a set of cuts that defines a first fiducial sample,
taking the limits as initially suggested by the previous completeness analysis.
However, as this solution might be an overly simplified way to create a complete
sample, for example, because it ignores spectroscopic follow-up in efficiency
for redshifts below $z_{lim}$, we also considered another set of cuts to define
a so-called conservative sample. This is smaller and therefore less
statistically constraining, but also even less prone to observational selection
effects. If the redshift drift is still significant in the conservative sample,
it would be even more meaningful in a carefully tailored selection-free sample.
These samples are adequate for the goal of this study, which is to develop a
first implementation of a model for drift in SN Ia properties. If fruitful, the
sample selection can later be refined as required with a more detailed model of
the observational selection, for instance, using the SNANA package
\citep{SNANA}.

The SNLS typically acquired SNe~Ia in the redshift range $0.4<z<0.8$; at these
redshifts, the rest-frame Bessel $B$ band roughly corresponds to the SNLS $i$
filter, which has a $5\sigma$ depth of
24.8~mag\footnote{\href{https://www.cfht.hawaii.edu/Science/CFHTLS/cfhtlsfinalreleaseexecsummary.html}{CFHT
final release website.}}. This converts into a $z_{\lim}=0.60$, in agreement
with \cite{neill2006}, \cite{perrett2010}, and \cite{bazin2011}. Fig.~14 of
\citet[][see their Section~5]{perrett2010}, however, suggests a lower limit of
$z_{\lim}=0.55$. We therefore used $z=0.60$ and $z=0.55$ as redshift limits for
the fiducial and conservative samples, respectively, for the SNLS.

Similarly, PS1 observed SNe~Ia in the range $0.2<z<0.4$, their $g$-band
$5\sigma$ depth is 23.1~mag \citep{rest2014}, which yields $z_{\lim}=0.31$, in
agreement with Fig.~6 of \cite{scolnic2018a}, for example. If we were to be
conservative, this figure would also suggest of a more stringent $z_{\lim}=0.27$
cut; we therefore used $0.31$ and $0.27$ for our fiducial and conservative
samples, respectively, for PS1.

In a similar redshift range, the SDSS has a limiting magnitude of 22.5
\citep{dilday2008, sako2008}, which would lead to $z_{\lim}=0.24$. However, the
SDSS surveys had to contend with limited spectroscopic resources. As discussed
in \citet[][Section~2]{kessler2009}, during the first year of SDSS, SNe~Ia with
$r<20.5$~mag were favored for spectroscopic follow-up, corresponding to a
redshift cut at $0.15$. For the remaining SDSS survey, additional spectroscopic
resources were available, and \cite{kessler2009} and \cite{dilday2008} showed a
reasonable completeness up to $z_{\lim}=0.2$. Based on this, we used
$z_{\lim}=0.20$ and $z_{\lim}=0.15$ for our fiducial and conservative samples,
respectively, for the SDSS.

\begin{figure}
    \centering
    \includegraphics[width=0.95\linewidth]{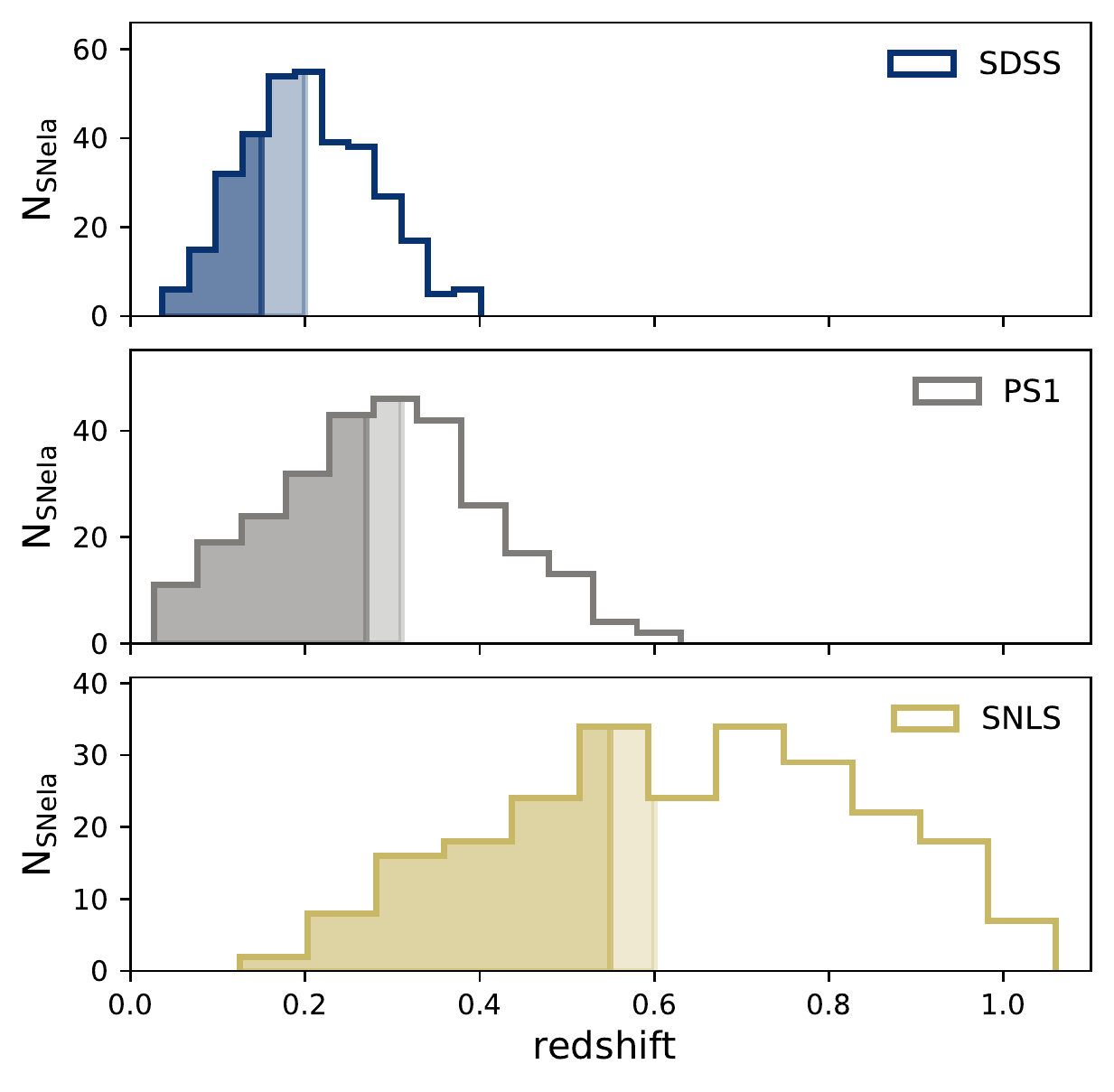}
    \caption{\textit{From top to bottom}: Redshift histograms of SNe~Ia from the
        SDSS, PS1, and SNLS dataset (data from Pantheon,
        \citealt{scolnic2018a}). The colored parts represent the distribution of
        SNe~Ia that we retained in our analysis because they are assumed to be
    free from observational selection bias (see Section~\ref{sec:sample}). The
darker (lighter) color responds to the conservative (fiducial) selection cut.}
    \label{fig:cuts}
\end{figure}

The sample selection is summarized in Table~\ref{tab:sample}, and the redshift
distribution of these three surveys is shown in Fig.~\ref{fig:cuts}. As
expected, the selected redshift limits are roughly located slightly before the
peak of these histograms. In Section \ref{ssec:verify} we confirm that these
redshift limits are effective for constructing nearly volume-limited subsamples
from samples that were initially more closely magnitude limited in their search
or spectroscopic follow-up.

\begin{table}
    \centering
    \caption{Composition of the SNe~Ia dataset used in this analysis.
    Conservative cuts are indicated in parentheses. The SNf limit is set
by \cite{rigault2020}, see text.}
    \label{tab:sample}
    \begin{tabular}{l c c}
        \hline\hline
        Survey & $z_{\lim}$ & $N_{\mathrm{SN}}$ \\
        \hline
        SNf & 0.08 & 114 \\
        SDSS & 0.20 (0.15) & 167 (82)\\
        PS1 & 0.31 (0.27) & 160 (122)\\
        SNLS & 0.60 (0.55) & 102 (78)\\
        HST & -- & 26 \\
        \hline
        Total & -- & 569 (422) \\
        \hline
    \end{tabular}
\end{table}

In addition, we used the SNe~Ia from the Nearby Supernova Factory
\citep[SNfactory,][]{aldering2002} published in \cite{rigault2020} and that have
been discovered from nontargeted searches (114 SNe~Ia, see their sections~3
and~4.2.2 ; see \citealt{aldering2020}). For this dataset, spectroscopic
screening was performed for candidates with $r \lesssim 19.5$; redshift cuts
were then applied when selecting which SN Ia to follow, resulting in a redshift
range of $0.02 < z < 0.09$, further reduced to $<0.08$ in \cite{rigault2020} to
extract local host properties. These 114 SNfactory SNe~Ia are thus in the
volume-limited part of the survey (Aldering et al., in prep.), and are therefore
assumed to be a random sampling of the underlying SN population. The SNfactory
sample is particularly useful for studying SN property drift because it enables
us to have a large complete SN~Ia sample at $z<0.1$. Finally, we include the HST
sample from Pantheon \citep{strolger04}. These high-redshift SNe are of great
interest as they provide the greatest leverage for testing evolution. While at
these redshifts the supernovae typing is challenging, the target classification
was robust enough to include them in the cosmological analysis
\citep{scolnic2018a}, and we did not impose further cuts.
Section~\ref{sec:results} highlights that while compatible with it, our results
are not dependent on the inclusion of this dataset.

\begin{figure}
    \centering
    \includegraphics[width=0.95\linewidth]{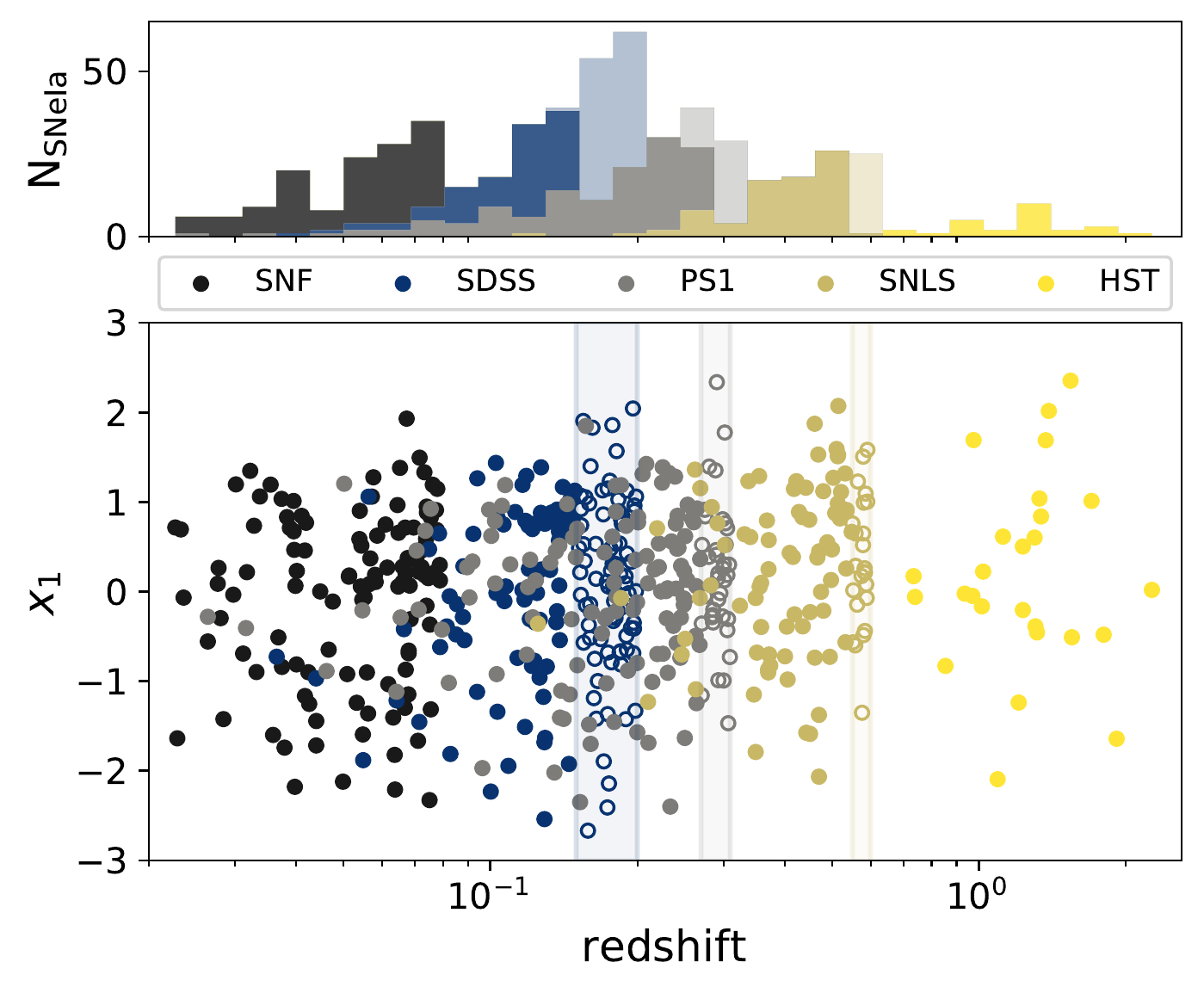}
    \caption{\textit{Bottom:} \textsc{\texttt{SALT2.4}} light-curve stretch as a
        function of redshift for each survey considered in this analysis (see
        legend). Solid (open) markers correspond to the conservative (fiducial)
        redshift cuts. \textit{Top:} Stacked redshift histograms in dark (light)
    colors for the conservative (fiducial) redshift cuts.}
    \label{fig:sample}
\end{figure}

We present the stretch distribution and redshift histogram of these five surveys
up to their respective $z_{\lim}$ in Fig.~\ref{fig:sample}. We observe here that
the fraction of low-stretch SNe (typically $x_1 < -1$) appears to decrease as a
function of redshift; this is confirmed in Fig.~\ref{fig:modelall}, in which the
evolution of the mean stretch is shown, with the data split in redshift bins of
regular sample size. SNe~Ia at higher redshift have a larger stretch ($0.34 \pm
0.10$ at $z\sim0.65$) on average than those at lower redshift ($-0.17\pm 0.10$
at $z\sim0.05$), suggesting that the underlying stretch distribution evolves
with redshift.

\subsection{Testing the construction of a volume-limited
sample}\label{ssec:verify}

In section~\ref{ssec:cuts} we have built volume-limited samples from a set of
magnitude-limited ones using simple redshift cuts. This simplified approach is
statistically suboptimal, but should suffice to test our key question whether
redshift evolution of stretch is compatible with the model of
\cite{rigault2020}. However, the possibility remains that a complex
observational selection function related to spectroscopic follow-up efficiencies
below our fiducial (or even conservative) redshift cuts might still affect our
sample, making it not fully volume limited; this would in turn bias our
conclusion about the astrophysical drift of the SNe~Ia population. We now
examine this possibility.

To test for the existence of potential leftover observational selection biases
in our sample, we compared the stretch and color distributions of the SNe~Ia
originating from different datasets having overlapping redshift ranges: these
distributions should be similar if they reflect the underlying parent
population. We note that the redshift range has to be narrow enough so that any
drift would be negligible.

\begin{figure}
    \centering
    \includegraphics[width=0.95\linewidth]{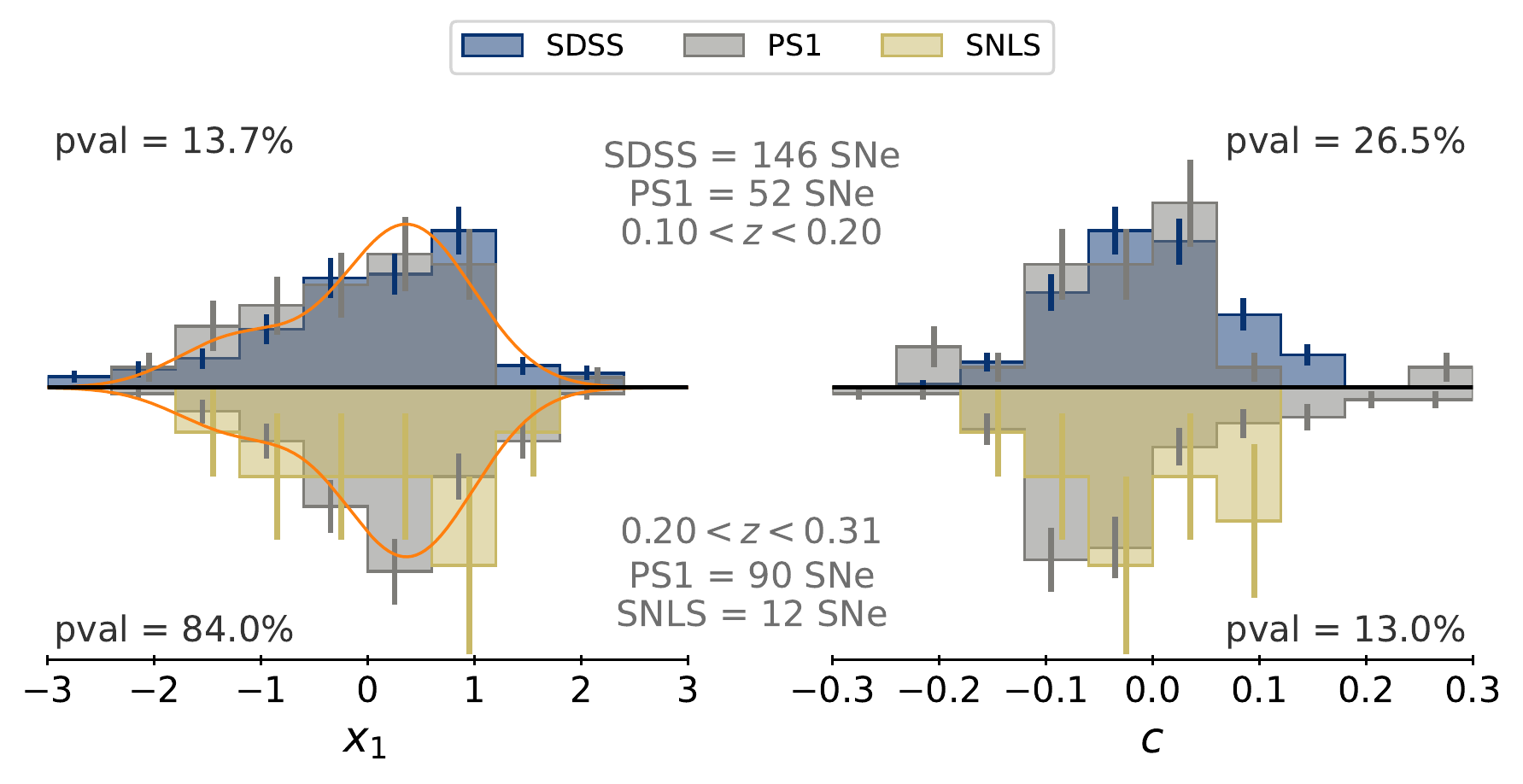}
    \caption{$x_1$ (left) and $c$ (right) distribution histograms of
            different surveys overlapping in redshift. \textit{Facing up}: SDSS
            and PS1 within the redshift range $0.10 < z < 0.20$. \textit{Facing
            down}: PS1 and SNLS within the redshift range $0.20 < z < 0.31$.
            Error bars show the Poisson noise. Our stretch base model is
            illustrated in orange at the mean redshift of the redshift ranges,
            $0.15$ and $0.25$, respectively. Kolmogorov-Smirnov test $p$-values
            are indicated at the top (bottom) of each panel and show no
            sign that the SDSS and PS1 (PS1 and SNLS) $x_1$ and $c$
            distributions are not drawn from the same underlying
    distributions.}
    \label{fig:distrib}
\end{figure}

The two samples that overlap most in redshift are PS1 and SDSS in the redshift
range $0.10 < z < 0.20$ (see Fig~\ref{fig:sample}). This overlapping subsample
consists of the 146 SNe~Ia at the high-redshift end of SDSS and thus is most
likely to be affected by residual observational selection effects (see the
corresponding discussion in section~\ref{ssec:cuts}). Over that same redshift
range, PS1 has 52 SNe~Ia that are in the lowest redshift bins and thus unlikely
to have any observational selection issue. To identify potential inconsistency
between the PS1 and SDSS subsamples, Fig.~\ref{fig:distrib} (upper panels)
compares the stretch and color distribution of these two surveys. The resulting
Kolmogorov-Smirnov (KS) similarity test $p$-values ($p >10\%$) do not support
any inconsistency, in agreement with the visual impression from
Fig.~\ref{fig:distrib}.

We performed a similar analysis for PS1 and SNLS over the redshift range $0.20 <
z < 0.31$ (Fig.~\ref{fig:distrib}, lower panels), where the same conclusion can
be drawn: There is no substantial sign of discrepancy in the stretch and color
distributions between the low and high end of our fiducial SNLS and PS1 samples,
respectively. Nonetheless, the small size of the SNLS dataset at $z < 0.31$ (12
SNe~Ia vs. 90 for PS1) limits the sensitivity of this test, and only a strong
deviation would be noticeable. Extending the redshift range to $0.20 < z < 0.40$
(although we have no PS1 data above 0.3) allows increasing the SNLS subsample to
31, but the stretch $p$-value remains high (34\%), showing no sign of
inconsistency.

We finally highlight that the SNe~Ia color is more prone to observational
selection effects than stretch, as illustrated in Fig.~\ref{fig:maglim}; see
also Fig.~3 of \cite{kessler2017}, for instance. Hence, because the comparison
of color distributions shows no significant indication of leftover observational
selection effect, this further supports our claim that our simple redshift-based
selection criteria are sufficient to build the complete SNe~Ia samples required
to test the redshift evolution of the stretch distribution.

\section{Modeling the redshift drift}\label{sec:modeling}

\cite{rigault2020} presented a model for the evolution of the fraction of
younger and older SNe~Ia as a function of redshift following previous work on
rates and delay-time distributions \citep[e.g.,][]{mannucci2005,
scannapieco2005, sullivan2006, smith2012, childress2014, maozmannucci2014}. In
short, it was assumed that the number of young SNe~Ia follows the star formation
rate (SFR) in the Universe, while the number of old SNe~Ia follows the number of
billion-old stars in the Universe, that is, the stellar mass (M$^*$). Hence, if
we denote $\delta(z)$ ($\psi(z) = 1-\delta(z)$) the fraction of young (old)
SNe~Ia in the Universe as a function of redshift, then the ratio $\delta/\psi$
is expected to follow the evolution of the specific star formation rate
(SFR/M$^*$), which goes as $(1+z)^{2.8}$ until $z\sim2$
\citep[e.g.,][]{tasca2015}. Since $\delta(0.05) \sim \psi(0.05)$
\citep{rigault2013, rigault2020, wiseman2020}, in agreement with rate
expectations \citep{mannucci2006, rodney2014}, \cite{rigault2020} concluded that
\begin{equation}
    \label{eq:delta}
    \delta(z) = \left( K^{-1} \times (1+z)^{-2.8} +1 \right)^{-1}
\end{equation}
with $K=0.87$. This model is comparable to the evolution subsequently
predicted by \cite{childress2014} based on SN rates in galaxies depending on
their quenching time as a function of their stellar mass.

\subsection{Base underlying stretch distribution}
\label{sec:basemodel}

\begin{figure*}
    \centering
    \includegraphics[width=0.8\linewidth]{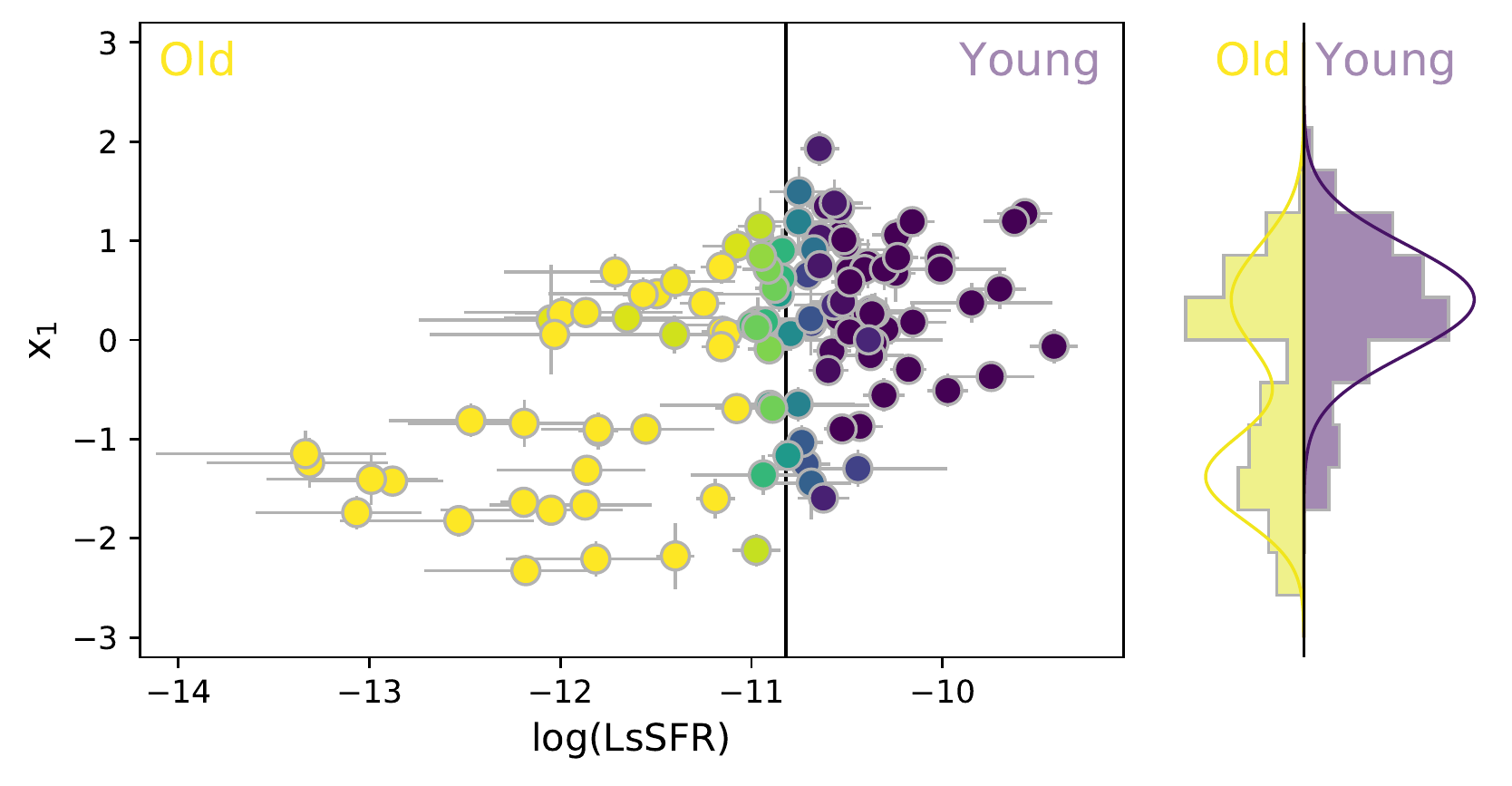}
    \caption{\textit{Main}: \textsc{\texttt{SALT2.4}} light-curve stretch
        ($x_1$) as a function of the LsSFR for SNfactory SNe. The color
        corresponds to the probability, $p_y$, for the SNe~Ia to be young, i.e.,
        to have $\log\mathrm{LsSFR} \geq -10.82$ \citep[see][]{rigault2020}.
    \textit{Right}: $p_y$-weighted histogram of the SN stretches, as well as the
adjusted base model; contributions of the younger and older population are shown
in purple and yellow, respectively.}
    \label{fig:stretchlssfr}
\end{figure*}

To model the evolution of the full SN stretch distribution as a function of
redshift, we need to model the SN stretch distribution for each age subsample
given our aforementioned model of the evolution of the fraction of younger and
older SNe~Ia with cosmic time. \cite{rigault2020} presented the relation between
SN stretch and LsSFR measurement, a progenitor age tracer, using the SNfactory
sample. This relation is shown in Fig.~\ref{fig:stretchlssfr} for the SNfactory
SNe used in the current analysis. Given the structure of the stretch-LsSFR
scatter plot, our model of the underlying SN~Ia stretch distribution is defined
as follows: the stretch distribution of the younger population
($\log(\mathrm{LsSFR})\geq-10.82$) is modeled as a single normal distribution
$\mathcal{N}(\mu_1, \sigma_1{}^2)$, and the stretch distribution of the older
population ($\log(\mathrm{LsSFR})<-10.82$) is modeled as a bimodal Gaussian
mixture $a\times \mathcal{N}(\mu_1, \sigma_1{}^2) + (1-a)\times
\mathcal{N}(\mu_2, \sigma_2{}^2)$, where one mode is the same as for the young
population, $a$ representing the relative effect of the two modes.

The stretch probability distribution function (pdf) of a given SN will be the
linear combination of the stretch distributions of these two population weighted
by its probability $y^i$ to be young (see Section~\ref{sec:basemodelapplied}).
In general, however, the fraction of young SNe~Ia as a function of redshift is
given by $\delta(z)$ (see Eq.~\ref{eq:delta}), and therefore our redshift drift
model of the underlying stretch distribution of SNe~Ia as a function of redshift
$X_1(z)$ is given by
\begin{align}\label{eq:stretchz}
    X_1(z) = \delta(z)&\times \mathcal{N}(\mu_1,\sigma_1{}^2) + \nonumber \\
    (1-\delta(z))&\times \left[ a\times\mathcal{N}(\mu_1,\sigma_1{}^2) +
    (1-a)\times\mathcal{N}(\mu_2,\sigma_2{}^2) \right]
\end{align}
This is our base drifting model.

\subsection{Comparison to data}\label{sec:basemodelapplied}

Given the probability $y^i$ that a given SN is young and assuming our base model
(see Section~\ref{sec:basemodel}), the probability of measuring a
\textsc{\texttt{SALT2.4}} stretch $x_1^i$ with an error $\mathrm{d}x_1^i$ is
given by
\begin{align}\label{eq:likelihoodsnf}
    \prob{x^i_1}{\vec{\theta}; \mathrm{d}x^i_1, y^i} =
    y^i & \times
    \mathcal{N}\left(x^i_1 \mid \mu_1, \sigma_1{}^2+\mathrm{d}x^i_1{}^2\right) +
    \nonumber\\
    (1-y^i) &\times \bigg[
    a \times \mathcal{N}\left(x^i_1 \mid \mu_1,
    \sigma_1{}^2+\mathrm{d}x^i_1{}^2\right) +
    \nonumber\\
    & (1-a) \times \mathcal{N}\left(x^i_1 \mid \mu_2,
    \sigma_2{}^{2}+\mathrm{d}x^i_1{}^2\right) \bigg].
\end{align}

The maximum-likelihood estimate of the five~free parameters
$\vec{\theta}\equiv({\mu_1, \mu_2, \sigma_1, \sigma_2,a})$ of the model is
obtained by minimizing the following:
\begin{equation}\label{eq:likelihood}
    -2\ln(L) = -2 \sum_i \ln \prob{x_1^i}{\vec{\theta};
    \mathrm{d}x_1^i, y^i}.
\end{equation}

Depending on whether $y^i$ can be estimated directly from LsSFR measurements,
there are two ways to proceed. We discuss them below.

\subsubsection{With LsSFR measurements}\label{sec:modelpy}

For the SNfactory sample, we can readily set $y^i = p^i_y$, the probability of
having $\log(\textrm{LsSFR}) \geq -10.82$ (see Fig.~\ref{fig:stretchlssfr}), to
minimize Eq.~\ref{eq:likelihood} with respect to $\vec{\theta}$. Results of
fitting the SNf SNe with this model are shown Table~\ref{tab:modelresults} and
illustrated in Fig.~\ref{fig:modelall}.

\begin{table*}
    \centering
    \caption{Best-fit values of the parameters for the base stretch distribution
    model when applied to the SNfactory dataset only (114 SNe~Ia), the fiducial
569 SN~Ia sample, or the conservative sample (422).}
    \label{tab:modelresults}
    \begin{tabular}{lccccc}
        \hline\hline
        Sample & $\mu_1$ & $\sigma_1$
               & $\mu_2$ & $\sigma_2$
               & $a$ \\
        \hline
        SNfactory & $ 0.41 \pm 0.08$ & $0.55 \pm 0.06$
                  & $-1.38 \pm 0.10$ & $0.44 \pm 0.08$
                  & $ 0.48 \pm 0.08$ \\
        Fiducial & $ 0.37 \pm 0.05$ & $0.61 \pm 0.04$
                 & $-1.22 \pm 0.16$ & $0.56 \pm 0.10$
                 & $ 0.51 \pm 0.09$ \\
        Conservative & $ 0.38 \pm 0.05$ & $0.60 \pm 0.04$
                     & $-1.26 \pm 0.13$ & $0.53 \pm 0.08$
                     & $ 0.47 \pm 0.09$ \\
        \hline
    \end{tabular}
\end{table*}

\subsubsection{Without LsSFR measurements}\label{sec:modelnopy}

When direct LsSFR measurements are lacking (i.e., $p_y^i$), we can extend the
analysis to non-SNfactory samples by using the redshift evolution of the
fraction $\delta(z)$ of young SNe~Ia (Eq.~\ref{eq:delta}) as a proxy for the
probability of a SN to be young. This still corresponds to minimizing
Eq.~\ref{eq:likelihood} with respect to the parameters
$\vec{\theta}\equiv(\mu_1, \mu_2, \sigma_1, \sigma_2, a)$ of the stretch
distribution $X_1$ (Eq.~\ref{eq:stretchz}), but this time, assuming $y^i =
\delta(z^i)$ for any given SN~$i$.\bigbreak

\noindent For the remaining analysis, we therefore minimized
Eq.~\ref{eq:likelihood} using $p_y^i$, the probability for the SN $i$ to be
young, when available (i.e., for SNfactory dataset), and $\delta(z^i)$, the
expected fraction of young SNe~Ia at the SN redshift $z^i$, otherwise.

\begin{figure*}
    \centering
    \includegraphics[width=0.7\linewidth]{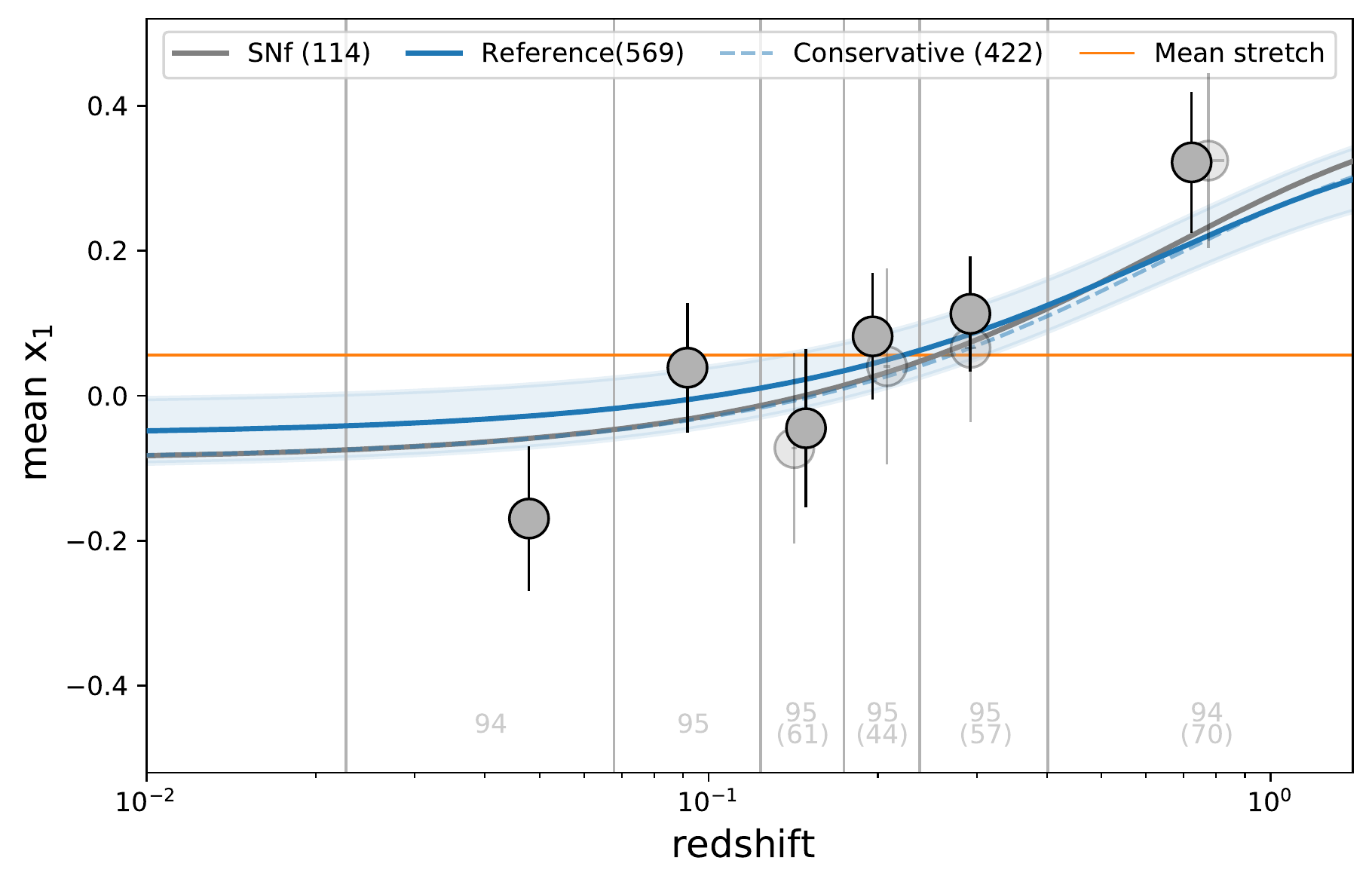}
    \caption{Evolution of the mean SN \textsc{\texttt{SALT2.4}} stretch ($x_1$)
        as a function of redshift. Markers show the stretch plain mean (the
        error is estimated from the scatter) measured in redshift bins of equal
        sample size, indicated in light gray at the bottom of each redshift bin.
        Full and light markers are used when the fiducial or conservative
        samples are considered, respectively. The horizontal orange line
        represents the mean stretch for the nonevolving Gaussian model (last
        line of Table~\ref{tab:comp}) on the fiducial sample. Best fits of our
        base drifting model are shown as blue, dashed blue, and gray when fitted
        on the fiducial sample, the conservative sample, or the SNfactory
        dataset only, respectively; all are compatible. The light blue band
        illustrates the amplitude of the error (including\ covariance) of the
        best-fit model when considering the fiducial dataset.}
    \label{fig:modelall}
\end{figure*}

Results of fitting this model to all the 569 (resp. 422) SNe from the fiducial
(conservative) sample are given Table~\ref{tab:modelresults}, and the predicted
redshift evolution of mean stretch (expected $x_1$ given the distribution of
Eq.~\ref{eq:stretchz}) illustrated as a blue band in Fig.~\ref{fig:modelall}
accounting for parameter errors and their covariances. This figure shows that
the measured mean SN~Ia stretch per redshift bin of equal sample size closely
follows our redshift drift modeling. This is indeed what is expected if old
environments favor low SN stretches \citep[e.g.,][]{howell2007} and if the
fraction of old SNe~Ia declines as a function of redshift. See
Section~\ref{sec:results} for a more quantitative discussion.

\subsection{Alternative models}\label{sec:othermodel}

In Section~\ref{sec:basemodel} we have modeled the underlying stretch
distribution following \cite{rigault2020}, that is,\ as a single Gaussian for
the young SNe~Ia and a mixture of two Gaussians for the old SNe Ia population,
one being the same as for the young population, plus another one for the
fast-declining SNe~Ia that appear to only exist in old local environments. This
is our so-called base model. However, to test different modeling choices, we
implemented a suite of alternative parameterizations that we also adjusted to
the data following the procedure described in Section~\ref{sec:modelnopy}. 

\cite{howell2007} used a simpler unimodal model per age category, assuming a
single normal distribution for each of the young and old populations. We thus
considered a Howell+drift model, with one single Gaussian per age group and the
$\delta(z)$ drift from Eq.~\ref{eq:delta}.

Alternatively, because we aim to probe the existence of an evolution with
redshift, we also tested constant models by restricting the base and Howell
models to use an assumed redshift-independent fraction $\delta(z) \equiv f$ of
young SNe; these models are hereafter labeled base+constant and Howell+constant.

We also considered another intrinsically nondrifting model, the functional form
developed for beams with bias correction \cite[BBC,][]{scolnic2016,
kessler2017}, used in recent SN cosmological analyses
\cite[e.g.,][]{scolnic2018a, descosmopaper2019, riess2016, riess2019} to account
for Malmquist biases. The BBC formalism assumes sample-based (hence
intrinsically nondrifting) asymmetric Gaussian stretch distributions:
$\mathcal{N}\left(\mu, \sigma_-{}^2\; \text{if} \;x_1<\mu,\; \text{else}
\;\sigma_+{}^2\right)$. The idea behind this sample-based approach is twofold:
(1) Malmquist biases are driven by survey properties, and (2) because current
surveys cover limited redshift ranges, having a sample-based approach covers
some potential redshift evolution information \citep{scolnic2016, scolnic2018a}.
See a more detailed discussion of BBC in Section~\ref{sec:discussion}. Finally,
for the sake of completeness, we also considered redshift-independent pure and
asymmetric Gaussian models. 

\section{Results}\label{sec:results}

\begin{figure}
    \centering
    \includegraphics[width=\linewidth]{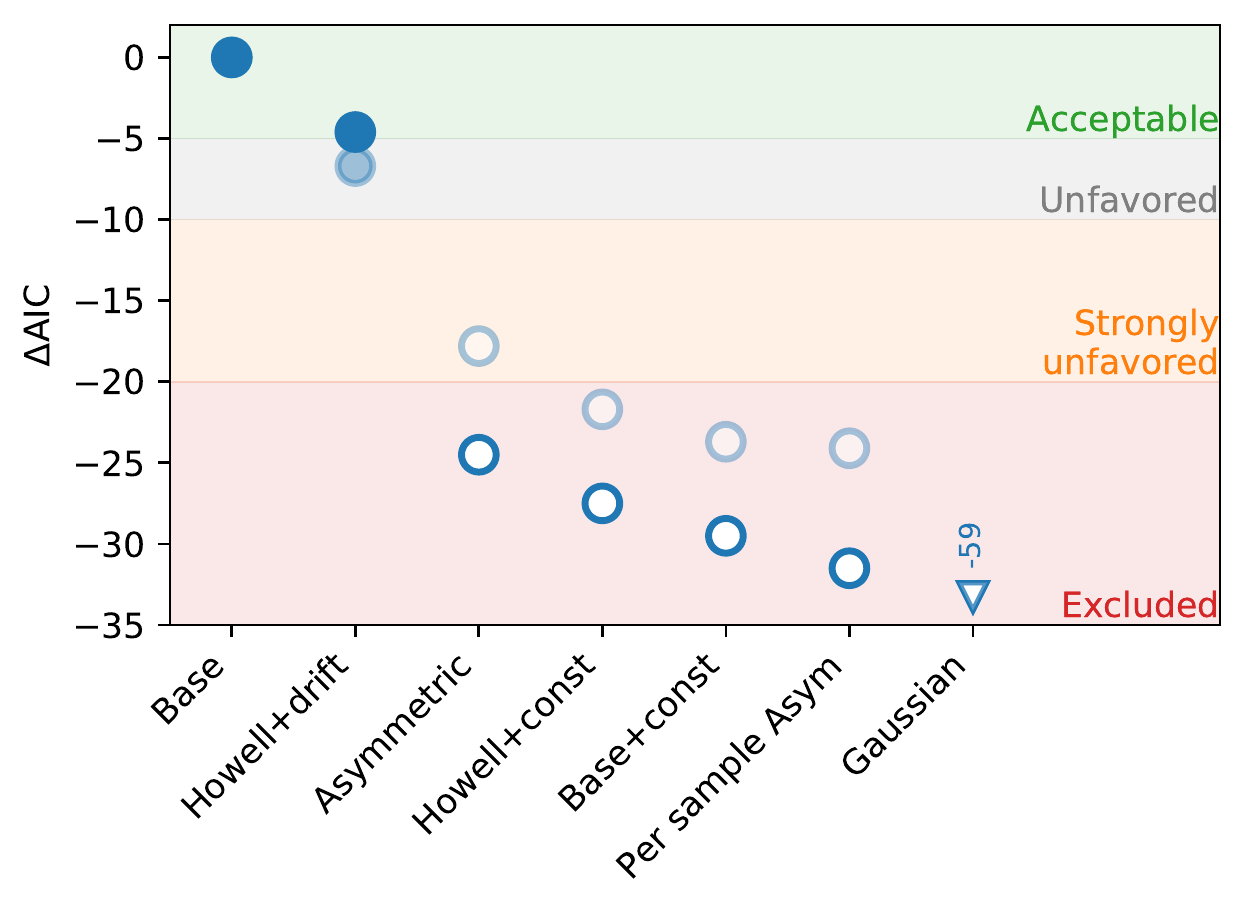}
    \caption{$\Delta$AIC between base model (reference) and other models (see
        Table~\ref{tab:comp}). Full and open blue markers correspond to models
        with and without redshift drift, respectively. Light markers show the
        results when the analysis is performed on the conservative sample rather
        than the fiducial one. Color bands illustrate the validity of the
        models, from acceptable ($\Delta\mathrm{AIC} > -5$) to excluded
        ($\Delta\mathrm{AIC} < -20$), see text. According to the AIC, all
    nondrifting models (open symbols) are excluded as a poorer representation of
the data than the base (drifting) model.}
    \label{fig:mod_comp}
\end{figure}

We adjusted each of the models described above on both the fiducial and
conservative samples (cf. Section~\ref{sec:sample}). The results are gathered in
Table~\ref{tab:comp} and are illustrated in Fig.~\ref{fig:mod_comp}. 

\begin{table*}
    \centering
    \caption{Comparison of the relative ability of each model to describe the
        data. For each considered model, we report whether the model is
        drifting, its number of free parameters, and for both the fiducial and
        the conservative cuts, $-2\ln(L)$ (see Eq.~\ref{eq:likelihood}), the AIC
        and the AIC difference ($\Delta$AIC) between this model and the base
    model used as reference because it has the lowest AIC.}
    \label{tab:comp}
    \begin{tabular}{ccc|ccc|ccc}
        \hline\hline
        & & & \multicolumn{3}{c}{Fiducial sample (569 SNe)}
            & \multicolumn{3}{|c}{Conservative sample (422 SNe)} \\
        Name & drift & $k$ &
        $-2\ln(L)$ & AIC & $\Delta$AIC & $-2\ln(L)$ & AIC & $\Delta$AIC\\
        \hline

        Base & $\delta(z)$ & 5
        & 1456.7 & 1466.7 & -- 
        & 1079.5 & 1089.5 & -- \\

        Howell+drift & $\delta(z)$ & 4
        & 1463.3 & 1471.3 & $-4.6$
        & 1088.2 & 1096.2 & $-6.7$ 
        \\

        Asymmetric & -- & 3
        & 1485.2 & 1491.2 & $-24.5$
        & 1101.3 & 1107.3 & $-17.8$ 
        \\

        Howell+const & $f$ & 5
        & 1484.2 & 1494.2 & $-27.5$
        & 1101.2 & 1111.2 & $-21.7$ 
        \\

        Base+const & $f$ & 6
        & 1484.2 & 1496.2 & $-29.5$
        & 1101.2 & 1113.2 & $-23.7$ 
        \\

        Per sample Asym. & per sample & 3$\times$5
        & 1468.2 & 1498.2 & $-31.5$
        & 1083.6 & 1113.6 & $-24.1$ 
        \\

        Gaussian & -- & 2
        & 1521.8 & 1525.8 & $-59.1$
        & 1142.6 & 1146.6 & $-57.1$ 
        \\
        \hline
    \end{tabular}
\end{table*}

Because the various models have different degrees of freedom, we used the Akaike
information criterion \citep[AIC, e.g.,][]{burnham2004} to compare their ability
to properly describe the observations. This estimator penalizes additional
degrees of freedom to avoid overfitting the data and is defined as follows:
\begin{equation}
    \mathrm{AIC} = -2\ln(L) + 2k,
\end{equation}
where $-2\ln(L)$ is derived by minimizing Eq.~\eqref{eq:likelihood}, and $k$ is
the number of free parameters to be adjusted. The reference model has the
smallest AIC; in comparison to this model, the models with
$\Delta\mathrm{AIC}<5$ are coined acceptable, those with
$5<\Delta\mathrm{AIC}<20$ are not favored, and those with
$\Delta\mathrm{AIC}>20$ are deemed excluded. This roughly corresponds to 2, 3,
and 5~$\sigma$ limits for a Gaussian probability distribution. 

The best model (with the smallest AIC) is the so-called base model and thus is
our reference model; this is true for the fiducial and conservative samples.
The base model also has the smallest $-2\ln(L)$, making it the most likely model
even when the overfitting issue that is accounted for by the AIC formalism is
ignored.

Furthermore, we find that redshift-independent stretch distributions are all
excluded as suitable descriptions of the data relative to the base model. The
best nondrifting model (the asymmetric model) has a very marginal chance ($p
\equiv \exp\left(\Delta\mathrm{AIC}/2\right) = 5\times10^{-6}$) to describe the
data as well as the base model. This result is just a quantitative assessment of
qualitative facts that are clearly visible in Fig.~\ref{fig:modelall}: The mean
SN stretch per bin of redshift strongly suggests a significant redshift
evolution rather than a constant value, and this evolution is well described by
Eq.~\ref{eq:delta}.

Surprisingly, the sample-based Gaussian asymmetric modeling used by current
implementations of the BBC technique \citep{scolnic2016, kessler2017} has one of
the highest AIC values in our analysis (see Section~\ref{sec:results}). While
its $-2\ln(L)$ is the smallest of all redshift-independent models (but still
$-11.5$ lower than the reference base model), it is strongly penalized for
requiring 15~free parameters ($\mu_0, \sigma_{\pm}$~for each of the five~samples
of the analysis). Hence, its $\Delta\mathrm{AIC}<-20$, which could be
interpreted as a probability $p=2\times 10^{-7}$ of being as good a
representation of the data as the base model.

We note that when models are compared that were adjusted on individual
subsamples rather than globally, the Bayesian information criterion
($\mathrm{BIC} = -2\ln(L) + k\ln(n)$, with $n$ the number of data points) might
be better suited than AIC because it explicitly accounts for the fact that each
subsample is fitted separately: the sample-based model BIC is rightfully the sum
of the BIC for each sample. We find $\Delta\mathrm{BIC}=-48$, again refuting the
sample-based asymmetric Gaussian model as being as pertinent as the base model.

In order to ensure that our results are not driven by the incompletely modeled
HST subsample, we recomputed $\Delta$AIC for each model excluding this dataset;
this did not change $\Delta$AIC by more than few tenths. The consistency of
these values with those in Table~\ref{tab:comp} shows that the HST subsample
does not drive our conclusions.

We report in Table~\ref{tab:bbc} our determination of $\mu_0$ and $\sigma_{\pm}$
for each sample when an asymmetric Gaussian model was implemented, and adjusted
on the nominally selection-free samples using our fiducial cuts (see
Section~\ref{sec:sample}). Our results are in close agreement with those of
\cite{scolnic2016} for the SNLS and SDSS and with the results reported by
\cite{scolnic2018a} for PS1, who derived these model parameters using the full
BBC formalism, using numerous simulations to model the observational selection
effects (see details, e.g., Section~3 of \citealt{kessler2017}). The agreement
between our fit of the asymmetric Gaussians on the supposedly selection-free
part of the samples and the results derived using the BBC formalism supports our
approach to constructing a sample with negligible observational selection
effects. If we were to use the \cite{scolnic2016} and \cite{scolnic2018a}
best-fit values of the $\mu_0, \sigma_{\pm}$ asymmetric parameters for the SNLS,
SDSS and PS1, respectively, the $\Delta$AIC between our base drifting model and
the BBC modeling would go even deeper from $-32$ to $-47$. We further discuss
the consequence of this result for cosmology in Section~\ref{sec:discussion}.
    
\begin{table}
    \centering
    \caption{Best-fit parameters for our sample-based asymmetric modeling of the
    underlying stretch distribution.}
    \label{tab:bbc}
    \begin{tabular}{ccccccc}
    \hline\hline
    Asymmetric & $\sigma_{-}$ & $\sigma_{+}$ & $\mu_0$ \\
    \hline
    SNfactory & 1.34 $\pm$ 0.13 & 0.41 $\pm$ 0.10 & 0.68 $\pm$ 0.15 \\
    SDSS & 1.31 $\pm$ 0.11 & 0.42 $\pm$ 0.09 & 0.72 $\pm$ 0.13 \\
    PS1 & 1.01 $\pm$ 0.11 & 0.52 $\pm$ 0.12 & 0.38 $\pm$ 0.16 \\
    SNLS & 1.41 $\pm$ 0.13 & 0.15 $\pm$ 0.13 & 1.22 $\pm$ 0.15 \\
    HST & 0.76 $\pm$ 0.36 & 0.79 $\pm$ 0.35 & 0.11 $\pm$ 0.44 \\
    \hline
    \end{tabular}
\end{table}
    
We also performed tests allowing the high-stretch mode of the old population to
differ from the young population mode, hence adding two degrees of freedom. The
corresponding fit is not significantly better, with a $\Delta$AIC of $-0.4$.
This reinforces our assumption that the young and old populations indeed appear
to share the same underlying high-stretch mode. Furthermore, we might wonder
whether a low-stretch mode might also exist in the young population (see
Fig.~\ref{fig:stretchlssfr}). We tested for this by allowing this population to
also be bimodal, finding the amplitude of such a low-stretch mode to be
compatible with~0 ($<2\%$) in this young population. More generally, this raises
the question of how well a given environmental tracer (here LsSFR) traces the
age. An analysis dedicated to this question will be presented in Briday et al.\
in prep.

Finally, ignoring the LsSFR measurements, which are only available for the
SNfactory dataset (see Section~\ref{sec:modeling}), reduces the significance of
the results presented in this section, as expected. Even so, nondrifting models
remain strongly disfavored. For instance, the best-fitting sample-based Gaussian
asymmetric model is still $\Delta\mathrm{AIC}<-10$, which is less representative
of the data than our base drifting model.

\section{Discussion}\label{sec:discussion}

To the best of our knowledge, a SN~Ia stretch redshift drift modeling has never
been explicitly used in cosmological analyses, although a Bayesian hierarchy
formalism such as UNITY \citep{rubin2015}, BAHAMAS \citep{shariff2016}, or Steve
\citep{hinton2019} can easily allow it (see, e.g., sections~1.3 and 2.5 of
\cite{rubin2015}). Not doing so is a second-order issue for SN cosmology because
it only affects the way in which the Malmquist bias is accounted for. As long as
the Phillips relation \citep{phillips1993} standardization parameter $\alpha$ is
not redshift dependent (a study that is beyond the scope of this paper, but see,
e.g., \citealt{scolnic2018a}), the stretch-corrected SNe~Ia magnitudes used for
cosmology are indeed blind to the underlying stretch distribution for complete
samples. However, surveys usually do have significant Malmquist bias for the
upper half of their SN redshift distribution. As a consequence, mismodeling of
the underlying stretch distribution will bias the SN magnitudes derived from
such surveys. 

Commonly used Malmquist bias correction techniques, such as the BBC-formalism,
assume per-sample asymmetric Gaussian functions to model the underlying stretch
and color distributions. As shown in Section~\ref{sec:results}, however, such a
sample-based distribution is excluded in comparison to our drifting model. In
contrast to what \citet[][Section~2]{scolnic2016} and
\citet[][Section~5.4]{scolnic2018a} have suggested, that is, that traditional
surveys span sufficiently limited redshift ranges such that the per-sample
approach accounts for implicit redshift drifts, a direct modeling of the
redshift drift is therefore more appropriate than a sample-based approach. We
add here that as measurements of modern surveys try to cover increasingly larger
redshift ranges in order to reduce calibration systematic uncertainties, this
sample-based approach becomes less valid, notably for PS1, DES and, soon, the
Large Synoptic Survey Telescope.

\begin{figure}
    \centering
    \includegraphics[width=\linewidth]{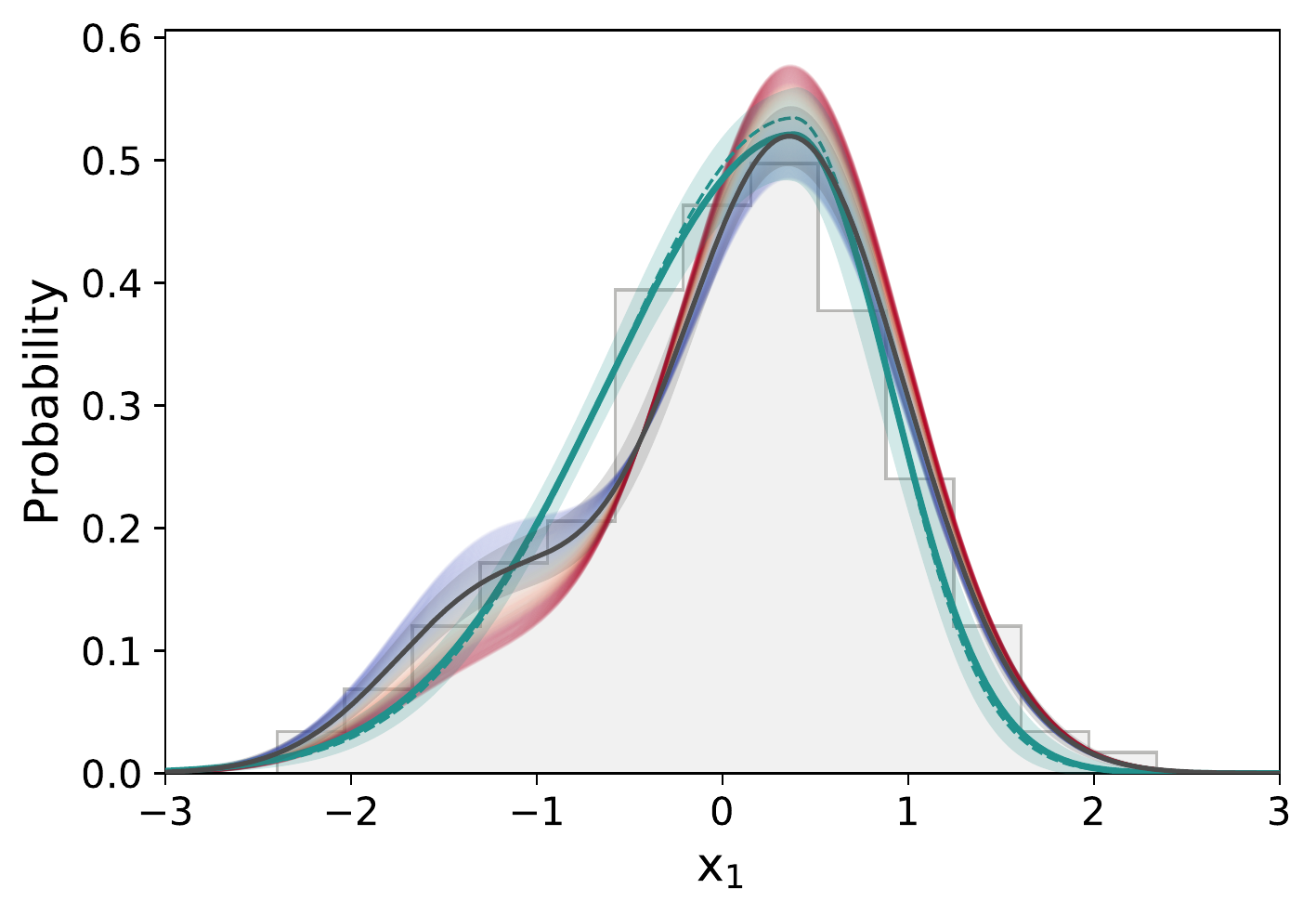}
    \caption{Distribution of the PS1 SN~Ia \textsc{\texttt{SALT2.4}} stretch
        ($x_1$) after the fiducial redshift limit cut (gray histogram). This
        distribution is supposed to be a random draw from the underlying stretch
        distribution. The green lines show the BBC model of this underlying
        distribution (asymmetric Gaussian). The full line (band) is our best fit
        (its error); the dashed line shows the \cite{scolnic2018a} result. The
        black line (band) shows our best-fit base modeling (its error, see
        Table \ref{tab:modelresults}) that includes redshift drift. For
        illustration, we show (colored from blue to red with increasing
        redshifts) the evolution of the underlying stretch distribution as a
        function of redshift for the redshift range covered by PS1 data.}
    \label{fig:bbc_pdf_ps1}
\end{figure}

We illustrate in Fig.~\ref{fig:bbc_pdf_ps1} the prediction difference in the
underlying stretch distribution between the per-sample asymmetric modeling and
our base drifting model for the PS1 sample. Our model is bimodal, and the
relative amplitude of each mode depends on the redshift-dependent fraction of
old and young SNe~Ia in the sample: the higher the fraction of old SNe~Ia (at
lower redshift), the higher the amplitude of the old-specific low-stretch mode.
This redshift dependence on the underlying stretch distributions is shown with
colors from blue to red in Fig.~\ref{fig:bbc_pdf_ps1} for the redshift range
covered by PS1. The observed $x_1$ histogram follows the model we defined using
the sum of individual underlying SN-redshift distributions. As expected, the two
modeling approaches differ mostly in the negative part of the SN stretch
distribution. The asymmetric Gaussian distribution goes through the middle of
the bimodal distribution, overestimating the number of SNe~Ia at $x_1\sim-0.7$
and underestimating it at $x_1\sim-1.7$ in comparison to our base drifting model
for typical PS1 SN redshifts. This means that the SN bias-corrected standardized
magnitude estimated at a redshift plagued by observational selection effects
would be biased by a mismodeling of the true underlying stretch distribution.

Assessing the amplitude of this magnitude bias for cosmology is beyond the scope
of this paper given the complexity of the BBC analysis. It would require a full
study using our base model (Eq.~\ref{eq:stretchz}) in place of the sample-based
asymmetric modeling as part of the BBC simulations. However, we already
highlighted that even if a nondrifting sample-based model could provide
comparable result in the volume-limited part of the various samples, these
models would differ when extrapolated to higher redshifts, precisely where the
underlying distribution will matter for correcting Malmquist biases.

In the era of modern cosmology, where we aim to measure $w_0$ at a subpercent
level and $w_a$ with 10\%\ precision \citep[e.g.,][]{lsstpaper}, we stress that
correct modeling of potential SN redshift drift should be further studied and
care should be taken when samples are used that are affected by observational
selection effects.

\section{Conclusion}\label{sec:ccl}

We have presented an initial study of the drift of the underlying SNe~Ia stretch
distribution as a function of redshift. We built effectively volume-limited
SN~Ia subsamples from the Pantheon dataset \citep[][SDSS, PS1, and
SNLS]{scolnic2018a}, to which we added HST and SNfactory data from
\cite{rigault2020} for the high- and low-redshift bins. We only considered the
SNe that have been discovered in the redshift range of each survey in which
observational selection effects are negligible, so that the observed SNe~Ia
stretches are a random sampling of the true underlying distribution. This
resulted in a fiducial sample of 569 SNe~Ia (422 SNe when more conservative cuts
were applied).

Following predictions made in \cite{rigault2020}, we introduced a redshift drift
model that depends on the expected fraction of young and old SNe~Ia as a
function of redshift, with each age population having its own underlying stretch
distribution.

In addition to this base model, we studied various distributions, including
redshift-independent models; we also studied the prediction from a per-sample
asymmetric Gaussian stretch distribution used, for instance, by the beams with
bias correction Malmquist bias correction algorithm \citep{scolnic2016,
kessler2017}. Our conclusions are listed below.
\begin{enumerate}

    \item The underlying SN~Ia stretch distribution is significantly redshift
        dependent, as previously suggested by~\cite{howell2007}, for example, in
        a way that observational selection effects alone cannot explain. This
        result is largely independent of the details of each age-population
        model.
    
    \item Redshift-independent models are quantitatively excluded as suitable
        descriptions of the data relative to our base model. This model assumes
        that (1) the younger population has a unimodal Gaussian stretch
        distribution while the older population stretch distribution is bimodal,
        one mode being the same as the young one, and (2) the evolution of the
        relative fraction of younger and older SNe~Ia follows the prediction
        made in \cite{rigault2020}. This second result further supports the
        existence of both young and old SN~Ia populations, in agreement with
        rate studies \citep{mannucci2005, scannapieco2005, sullivan2006,
        aubourg2008}. 
        
    \item Models using survey-based asymmetric Gaussian distributions, for
        instance, as employed in the current implementation of BBC, are excluded
        as a good description of the data relative to our drifting model. This
        means that the sample-based approach does not accurately account for
        redshift drift, a problem that will be exacerbated as surveys span
        increasingly larger redshift ranges. As a result, even if the necessary
        extra degrees of freedom might be acceptable given the large number of
        SNe~Ia in cosmological studies, extrapolating the SN property
        distributions from the volume-limited part of a survey to its
        Malmquist-biased magnitude-limited part would still be inaccurate
        because of the redshift evolution.

    \item Given the current dataset, we suggest the use of the following stretch
        population model as a function of redshift:
        \begin{align*}
        \label{eqconclusion:stretchz}
            X_1\left(z \right) =
            \delta(z)&\times\mathcal{N}(\mu_1,\sigma_1{}^2)\,+\nonumber\\
            (1-\delta(z))&\times \left[a\times\mathcal{N}(\mu_1,\sigma_1{}^2) +
            (1-a)\times\mathcal{N}(\mu_2,\sigma_2{}^2)\right],
            \tag{\ref{eq:stretchz}}
        \end{align*}
        with $a=0.51$, $\mu_1=0.37$, $\mu_2=-1.22$, $\sigma_1=0.61$, and
        $\sigma_2=0.56$ (see Table~\ref{tab:modelresults}), and using the
        age-population drift model, \begin{align*}
            \delta(z) & = \left( K^{-1} \times (1+z)^{-2.8} +1 \right)^{-1}
            \tag{\ref{eq:delta}}
        \end{align*}
        with $K=0.87$.
\end{enumerate}

We considered a simple two-population Gaussian mixture modeling. Additional data
free from significant Malmquist bias would enable us to refine it as required.
We note that samples at the low- and high-redshift ends of the Hubble diagram
would be particularly helpful for this drifting analysis; fortunately, this will
soon be provided by the Zwicky Transient Facility \citep[low-$z$,][]{bellm2019,
graham2019} and the Subaru and SeeChange SNe~Ia programs (high-$z$),
respectively. 

The next step in this line of analysis will incorporate our model into the SNANA
framework \citep{SNANA}, both to more accurately account for observational
selection functions and to test the effect of our model on the derivation of
cosmological parameters. This study is currently being conducted.

\begin{acknowledgements}
    This project has received funding from the European Research Council (ERC)
    under the European Union's Horizon 2020 Research and Innovation program
    (grant agreement no 759194 - USNAC).
    This work was supported in part by the Director, Office of Science, Office
    of High Energy Physics of the U.S. Department of Energy under Contract No.
    DE-AC025CH11231.
    This project is partly financially supported by Région Rhône-Alpes-Auvergne.
\end{acknowledgements}

\bibliographystyle{aa}

\end{document}